%% file: main.tex
\documentclass[copyright,creativecommons]{eptcs}

\providecommand{\event}{QAPL 2010} 
\providecommand{\volume}{??} \providecommand{\anno}{20??}
\providecommand{\firstpage}{1} \providecommand{\eid}{??}
\usepackage{breakurl}        
\usepackage[all]{xy}
\usepackage{color,graphicx}
\usepackage{cite}
\usepackage[fleqn]{amsmath}
\usepackage{amssymb}
\usepackage{amscd}
\usepackage{stmaryrd}
\usepackage{hyperref}
\usepackage{version}
\usepackage{theorem}
\usepackage{subfig}
\usepackage{wrapfig}
\usepackage{marvosym}
\usepackage{wasysym}

\input{preamb}

\title{Testing Reactive Probabilistic Processes}
\author{Sonja Georgievska and Suzana Andova
\institute{Department of Mathematics and Computer Science \\ Eindhoven University of Technology\\ The Netherlands}\\
 \email{s.georgievska@tue.nl, s.andova@tue.nl}}
\def\titlerunning{Testing Reactive Probabilistic Processes}
\def\authorrunning{Sonja Georgievska and Suzana Andova}

\excludeversion{LONGVERSION}

\begin{document}
\maketitle

\begin{abstract}
We define a testing equivalence in the spirit of De Nicola and
Hennessy for reactive probabilistic processes, i.e. for processes
where the internal nondeterminism is due to random behaviour.
We characterize the testing equivalence in terms of ready-traces.
From the characterization it follows that the equivalence is
insensitive to the exact moment in time in which an internal
probabilistic choice occurs, which is inherent from the original
testing equivalence of De Nicola and Hennessy.  We also show
decidability of the testing equivalence for finite systems for which
the complete model may not be known.
\end{abstract}

\input{introduction1}

\input{prelim}
\input{ready}

\input{testing_def}

\input{relationship}
\input{decidability}

\input{conclusion}
\bibliographystyle{eptcs}
\bibliography{testingZ}{}
\end{document}

%% file: preamb.tex
\newcommand{\actset}{\mathcal{A}}

\newcommand{\exec}[1]{\mathrel{\xrightarrow{#1}}}

\newcommand{\nexec}[1]{\mathalpha{\not\exec{#1}}}
\newcommand{\pexec}[1]{\mathrel{\stackrel{#1}{\dashrightarrow}}}

\newcommand{\tick}{\mapsto}
\newcommand{\ntick}{\mathalpha{\not\mapsto}}
\newcommand{\success}{\omega}

\newcommand{\terma}[1]{\exec{a}\check}
\newcommand{\nterma}[1]{\nexec\check}
\newcommand{\termt}[1]{\tick\check}
\newcommand{\ntermt}[1]{\ntick\check}

\def \lparal{\mathbin{\setbox0=\hbox{$\|$}
    \dimen0=\dp0 \advance\dimen0 -1.5pt \dp0=\dimen0
    \underline{\kern-1.5pt\box0\kern1.5pt}}}

\newcommand{\sosaltact}{\mathalpha{\ensuremath{}}}

\newcommand{\sosrepact}{\mathalpha{\ensuremath{}}}
\newcommand{\sosparactone}{\mathalpha{\ensuremath{}}}
\newcommand{\sosparacttwo}{\mathalpha{\ensuremath{}}}

\newcommand{\sosencterm}{\mathalpha{\ensuremath{}}}
\newcommand{\sosencact}{\mathalpha{\ensuremath{}}}

\newcommand{\soshideterm}{\mathalpha{\ensuremath{}}}
\newcommand{\soshideactone}{\mathalpha{\ensuremath{\langle\textsf{hide-act}_1\rangle}}}
\newcommand{\soshideacttwo}{\mathalpha{\ensuremath{\langle\textsf{hide-act}_2 \rangle}}}

\newcounter{sosrule}
\newcommand{\sys}{\mathcal{P}}
\newcommand{\system}{(\nstates, \pstates, \atransym, \ptransym)}

\newcommand{\prob}{P}

  \newcommand{\bisimilar}[1][]{%
     \setbox0=\hbox{\kern-.1ex{$\leftrightarrow$}\kern-.1ex}
     \setbox1=\vbox{\hbox{\raise .1ex \box0}\hrule}%
     \ensuremath{\mathrel{\hbox{\kern.1ex\box1\kern.1ex}_{#1}}}
   }

\newcommand{\ptran}[1]{\pexec{#1}}

\newcommand{\ptransym}{\mathalpha{\dashrightarrow}}
\newcommand{\pitran}[1]{\ptran{#1}}

\newcommand{\atran}[1]{\exec{#1}}
\newcommand{\atransym}{\mathalpha{\rightarrow}}

\newcommand{\nstates}{S_n}
\newcommand{\pstates}{S_p}
\newcommand{\states}{S}

\usepackage[all]{xy}
\xyoption{color} \xyoption{arc} \xyoption{matrix}\xyoption{frame}
\xyoption{dvips}
\newxyColor{mygray1}{0.7}{gray}{}
\newxyColor{mygray2}{0.9}{gray}{}
\newxyColor{mygray3}{0.5}{gray}{}
\newxyColor{mygray4}{0.0}{gray}{}
\newxyColor{mygray5}{0.3}{gray}{}
\newxyColor{mygray6}{0.7}{gray}{}
\newxyColor{mygray7}{0.45}{gray}{}

\newdir{...}{{}*{}*{}}
\newcommand{\slabl}[1]{\ar@(r,u)@{...}^{{#1\hspace{1.1mm}}}}
\newcommand{\slabr}[1]{\ar@(l,u)@{...}_{\hspace{1.1mm}#1}}

\newcommand{\shorten}[1]{}
\newcommand{\Result}{\mathsf{R}}
\newcommand{\tests}{\mathcal{T}}
\newcommand{\menu}{M}
\newcommand{\obs}{\mathcal{O}}

\newcommand{\extch}{\mathsf{\sum}}

\newcommand{\barbed}{\mathrel{\approx_{\obs}}}
\newcommand{\testing}{\mathrel{\approx_{\mathcal{T}}}}

\newcommand{\pst}{\circ}
\newcommand{\nst}{\circ}
\newcommand{\xmenu}{X}

\newcommand{\ratexpr}{\mathcal{R}}

\newcommand{\poa}{\parallel}
\newcommand{\Det}{\mathsf{Det}}

\newtheorem{theorem}{Theorem}[section]
\newtheorem{lemma}[theorem]{Lemma}
\newtheorem{proposition}[theorem]{Proposition}

\newtheorem{definition}[theorem]{Definition}

\newenvironment{proof}[1][Proof]{\begin{trivlist}
\item[\hskip \labelsep {\bfseries #1}]}{\end{trivlist}}
\newenvironment{example}[1][Example]{\begin{trivlist}
\item[\hskip \labelsep {\bfseries #1}]}{\end{trivlist}}
\newenvironment{remark}[1][Remark]{\begin{trivlist}
\item[\hskip \labelsep {\bfseries #1}]}{\end{trivlist}}

\newcommand{\qed}{\nobreak \ifvmode \relax \else
      \ifdim\lastskip<1.5em \hskip-\lastskip
      \hskip1.5em plus0em minus0.5em \fi \nobreak
      \vrule height0.75em width0.5em depth0.25em\fi}

%% file: introduction1.tex
\section{Introduction}\label{sec:introduction}

A central paradigm behind process semantics based on observability
(e.g.~\cite{csp}) is that the exact moment an internal
nondeterministic choice is resolved is unobservable. This is because
an observer does not have insight into the internal structure of a
process but only in its externally visible actions. Unobservability
of  internal choice has been also achieved by the testing
theory~\cite{DH84,Hen88}\footnote{As shown in~\cite{D87} the process semantics based on~\cite{csp} and~\cite{DH84} coincide for a broad class of processes.}, where two processes are treated
equivalent iff they can not be distinguished when interacting with
their environment (which is an arbitrary process itself).
 It is natural, therefore, for this property to hold when internal choice
is quantified with probabilities. It turned out, however, that it was
not trivial to achieve unobservability of internal probabilistic
choice in probabilistic testing theory. The following example illustrates some points that cause this problem.

\begin{figure}[t]\centering
\large
$\xymatrix@R=0.3cm@C=0.01cm{
    & & & s \\
    & & & \nst  \ar@{-->}[dll]_{\frac{1}{2}} \ar@{-->}[drr]^{\frac{1}{2}} \\
& \nst \ar[dl]_{h} \ar[dr]^{t} & & & & \nst  \ar[dl]_{h} \ar[dr]^{t} \\ 
          \nst \ar[d]^{p}  && \nst   & & \nst&& \nst\ar[d]_{p} \\ 
           \nst &&  & &  && \nst  \\ 
           }$
         \quad
$\xymatrix@R=0.3cm@C=0.01cm{ u \\
   \nst \ar@/_/[d]_{h} \ar@/^/[d]^{t} \\ 
          \nst \ar[d]_{p} \\  
           \nst\ar[d]_{\smiley} \\ 
            \nst \\ 
           }$
           \quad
           $\xymatrix@R=0.3cm@C=0.01cm{
&&& \small s\|u \large \\
     & & & \nst  \ar@{-->}[dll]_{\frac{1}{2}} \ar@{-->}[drr]^{\frac{1}{2}} \\
& \nst \ar[dl]_{} \ar[dr]^{} & & & & \nst  \ar[dl]_{} \ar[dr]^{} \\ 
          \nst \ar[d]_{}  && \nst   & & \nst&& \nst\ar[d]_{} \\ 
           \nst\ar[d]_{\smiley} &&  & &  && \nst\ar[d]_{\smiley}  \\ 
           \nst &&  & &  && \nst  \\
           }$ \quad
         $\xymatrix@R=0.3cm@C=0.01cm{
         & & & \bar{s} \\
    & & & \nst  \ar[dll]_{h} \ar[drr]^{t} \\
& \pst \ar@{-->}[dl]_{\frac{1}{2}} \ar@{-->}[dr]^{\frac{1}{2}} & & & & \pst  \ar@{-->}[dl]_{\frac{1}{2}} \ar@{-->}[dr]^{\frac{1}{2}} \\ 
          \nst\ar[d]^{p}  && \nst  & & \nst&& \nst\ar[d]_{p} \\ 
          \nst &&  & &  && \nst   
           }$
 \caption{\small Processes $s$ and $\bar{s}$ are distinguished in  probabilistic may/must testing theory }
 \hbox{}
 \label{fig_machine}
\end{figure}
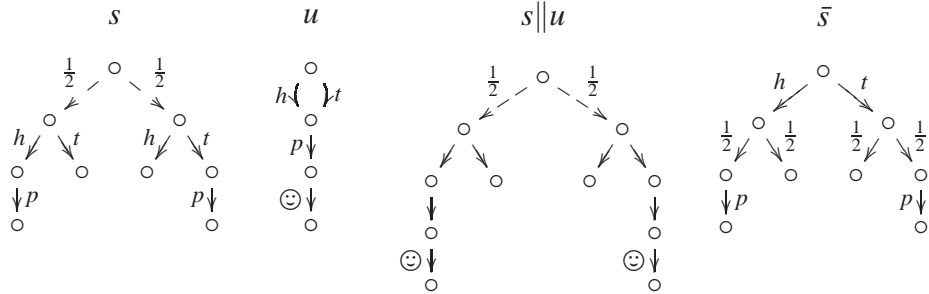

Consider a system consisting of a machine and a user, that
communicate via
 a menu of two buttons ``head'' and ``tail''  positioned at the machine.
  The machine makes a fair choice whether to give a prize if
  ``head'' is chosen or if ``tail'' is chosen. The user can choose ``head'' or ``tail'' by pressing the appropriate button. If the  user chooses the right outcome, a prize follows.    Note that by no means the machine's choice could have been revealed  beforehand to the user.
The machine can be modeled by the process graph
 $s$ in Fig.~\ref{fig_machine}. That is,
in half of the machine runs, it offers a prize after the ``head''
button has been pressed (out of the two-button menu ``head'' and
``tail''),
 while in the other half of the runs it offer a prize after the ``tail''
 button has been pressed (out of the two-button menu ``head'' and ``tail'').
 The user can be modeled by process $u$ in Fig.~\ref{fig_machine}.
 Sometimes she would press ``head'' and sometimes ``tail''; however,
 her goal is to win a prize, denoted by action $p$, and be ``happy'' afterwards, denoted by action  $\smiley$.

Let the user and the machine interact, i.e. let them synchronize on
all actions, except on the ``user happiness'' reporting action
$\smiley$. In terms of testing theory~\cite{DH84},   process $s$ is
tested with test $u$.
It can be computed, by means of the probability theory, that the probability with which the user has guessed the  machine choice is $\frac{1}{2}$. That is, the probability of a  $\smiley$ action being reported is $\frac{1}{2}$. However,  most  of the existing approaches for probabilistic testing, in particular probabilistic may/must testing~\cite{YL92,JY02,segala_testing96,DGHM08, PNM07},
   do not give this answer.
In order to compute the probability of $\smiley$ being reported, the
approaches in~\cite{YL92,JY02,segala_testing96,DGHM08, PNM07} use
\emph{schedulers} to resolve the action choice.  These schedulers
are taken very general and they are given the power to have insight
into the internal structure of the synchronized process.
Consider the synchronization $s \poa u$ represented by the graph in
Fig.~\ref{fig_machine},  where actions are hidden after they have
synchronized. A scheduler resolves the choices of actions in the two
states reachable in one probabilistic step from the initial state of
the graph $s \poa u$, thus yielding a fully probabilistic system.
For $s \mathrel{\poa} u$ in Fig.~\ref{fig_machine}, there are four
possible schedulers. They yield the following set of probabilities
with which $s$ passes the test $u$: $\{0, \frac{1}{2}, 1\}$. We can
see that, because the power of the schedulers is unrestricted,
nonviable upper and lower bounds for the probability are obtained.
Observe that this happens due to the effect of ``cloning'' the
action choice of $h$ and $t$ (the choice between $h$ and $t$ has
been ``cloned'' in both futures after the probabilistic choice in $s
\poa u$), and allowing a scheduler to schedule differently in the
two ``clones''. This, in fact, corresponds to a model where the user
is given 
power to \emph{see} the result of the probabilistic choice made by the machine \emph{before} she makes her guess. However, this is not the model we had initially in mind when the separate components, the machine and the user,  were specified.

 Consider now  process
$\bar{s}$ in Fig.~\ref{fig_machine}. To the user this process may as
well represent the behaviour of the machine -- the user cannot see
whether the machine makes the choice \emph{before} or \emph{after}
making the ``head or tail'' offers. According to the user, the
machine acts as specified as long as she is able to guess the result
in half of the cases. In fact, both schedulers, obtained by methods
in~ \cite{YL92,JY02,segala_testing96,DGHM08,PNM07}, when applied to
$\bar{s} \poa u$  yield exactly probability $\frac{1}{2}$ of
reporting action $\smiley$. Consequently, none of the approaches
in~\cite{YL92,JY02,segala_testing96,DGHM08,PNM07} equate processes
$s$ and $\bar{s}$: when tested with test $u$, they produce different
bounds for the probabilities of reporting $\smiley$. On the other
hand, if the probabilities are ignored and the probabilistic choice
is treated as an internal choice, processes $s$ and $\bar{s}$ are
equivalent by the  testing equivalence of~\cite{DH84}.

Being able to equate $s$ and $\bar{s}$ means allowing distribution
of external choice over internal probabilistic choice~\cite{csp}.
%
Actually, distribution of external choice over internal choice is
closely related to distribution of action prefix over internal
choice.
If distribution of external choice over internal probabilistic
choice is not allowed, then distribution of action prefix over
internal probabilistic choice  is questioned too, otherwise the
congruence properties of  \emph{asynchronous} or \emph{concurrent}
parallel composition~\cite{csp} (where processes synchronize on
their common actions while  interleave on the other actions) would
not hold. For instance, we would not be able to equate processes
$e.a. (b \oplus_{\frac{1}{2}}c)$ and $e.((a.b)
\oplus_{\frac{1}{2}}(a.c))$. (The operator ``.'' stands for
prefixing and the operator ``$\oplus$'' stands for a probabilistic
choice.) Running each of these two processes concurrently with
process $e.d$, yield processes that, unless distribution of external
choice over internal probabilistic choice is allowed, cannot be
equated. If we are not able to relate processes that differ only
 in the moment internal probabilistic choice is resolved,
 before or after an action execution (in other words, if we do not allow distribution of action prefix
over internal probabilistic choice), then for verification
 we can only rely on equivalences that inspect the internal
structure of processes, as bisimulations and
simulations~\cite{spectrum1}.

Motivated by the previous observations, in~\cite{fossacs10} we
propose a testing preorder which can deal with this problem.
According to this testing semantics, the probability with which
process $s$ passes test $u$ (Fig.~\ref{fig_machine}) is exactly
$\frac{1}{2}$.  The model considered in~\cite{fossacs10} is rather
general and allows probabilistic as well as internal
non-deterministic choice, in addition to action choice. Moreover,
the testing preorder is given a characterization in terms of a
probabilistic ready-trace preorder (a ready-trace is an alternating
sequence of ``action menus'' and executed actions). From this
characterization it follows that the underlying equivalence equates
processes $s$ and $\bar{s}$.

Since the tests in the model of \cite{fossacs10} have internal
transitions, in general, infinitely many tests need to be considered
to determine equivalence between two finite processes. Therefore,
the decidability of the testing equivalence for the general model at
the moment relies on the characterization of the equivalence in
terms of ready-traces. However, in practice, if we aim at  testing
whether the \emph{system} is equivalent to the \emph{model}, we may
not have access to the ready-traces and the internal transitions of
the system that are necessary to establish the equivalence. It is,
therefore, of practical interest to investigate for which type of
systems there exists a procedure to decide testing equivalence based
only on testing itself (see also \cite{verena_testing, LNR06, DLZ06}
for similar discussions).

In this paper we investigate decidability for systems of the testing
equivalence of~\cite{fossacs10} for \emph{reactive probabilistic
systems}~\cite{LS91}, where \emph{all} internal nondeterminism is
due to random behaviour. We first point out that, under the
condition that a test ``knows'' the current set of actions on which
it can synchronize with the system (i.e. the menu of
actions-candidates for synchronization), there exists a statistical
procedure to estimate the result of testing a system with a given
test. We then show that the set of tests necessary to determine
equivalence of two finite systems is finite, from which the
decidability result follows directly.

More concretely, we prove that deterministic (i.e.
non-probabilistic) tests suffice for distinguishing between finite
processes. This result follows from the proof that the testing
equivalence coincides with the probabilistic ready-trace
equivalence. In this paper we also present the characterization
proof, which is technically  much more involved than the
corresponding proof in \cite{fossacs10}, due to tests having ``less
power'' than in \cite{fossacs10}. From this characterization it also
follows that the testing equivalence, when applied to the model of
reactive probabilistic processes, preserves the previously mentioned
desirable properties:  it is insensitive to the exact moment of
occurrence of an internal probabilistic choice and it refines the
equivalence for the non-probabilistic case proposed in \cite{DH84}.

\paragraph{Structure of the paper} In Sec. \ref{sec:prelim} we
define some notions needed for the rest of the paper. In Sec.
\ref{sec:barbed} we recall the definition of probabilistic ready
trace equivalence from \cite{fossacs10}. In Sec. \ref{sec:testing}
we define a testing equivalence for the reactive probabilistic
processes. In Sec. \ref{sec:relationship} we prove that the
equivalences defined in sections \ref{sec:barbed} and
\ref{sec:testing} coincide. In Sec. \ref{sec:decidability} we show
the decidability results for the testing equivalence. Sec.
\ref{sec:conclusion} ends with discussion of related work, other
than \cite{fossacs10}, and concluding remarks.

%% file: prelim.tex
\section{Preliminaries}\label{sec:prelim}
We define some preliminary notions needed for the rest of the paper.
\subsection{Bayesian probability} For a set $A$, $2^A$ denotes its
power-set.
The following definitions are taken from~\cite{lindley}.

We consider a sample space, $\Omega$, consisting of points called
\emph{elementary events}. Selection of a particular $a \in \Omega$
is referred to as an ``$a$ has occurred''. An \emph{event} is a set
of elementary events. $A, B, C$ range over events. An event $A$
\emph{has occurred} iff, for some  $a \in A$, $a$ has occurred. Let
$A_1, A_2, \ldots$ be a sequence of events and $C$ be an event. The
members of the sequence are \emph{exclusive given C}, if whenever
$C$ has occurred no two of them can occur together, that is, if $A_i
\cap A_j \cap C=\emptyset$ whenever $i\not = j$. $C$ is called a
\emph{conditioning} event. If the conditioning event is $\Omega$,
then ``given $\Omega$'' is  omitted.

 For certain pairs of events $A$ and $B$, a real number $P(A|B)$ is
 defined and called the \emph{probability} of $A$ given $B$. These
 numbers satisfy the following axioms:

 \begin{enumerate}
 \item[A1:] $0 \leq P(A|B)\leq1$ and $P(A|A)=1$.
 \item[A2:]  If the events in $\{A_i\}_{i=1}^{\infty}$ are
 exclusive given $B$, then  $P(\cup_{i=1}^{\infty}
 A_i\ |\ B)=\sum_{i=1}^{\infty}P(A_i | B).$
 \item[A3:]  $P(C|A \cap B)\cdot P(A|B)=P(A \cap C|B)$.
 \end{enumerate}
For $P(A|\Omega)$ we simply write $P(A)$.

 \subsection{Probabilistic transition systems} In a probabilistic transition system (PTS)  there are two types of  transitions,
 viz. action and probabilistic transitions;
 a state can either perform action transitions only
(action state) or (unobservable) probabilistic transitions
only (probabilistic state). To simplify, we assume that
probabilistic transitions lead to action states. In
action states
the choice is between a set of actions,
but once the action is chosen, the next  state is determined. The
outgoing transitions of a probabilistic state $s$ define probability
over the power-set of the set of action states.

We give a formal definition of a PTS. Presuppose a finite set of
actions $\actset$.

\begin{definition}[Probabilistic Transition System (PTS)]
\label{def:PTS} A \emph{PTS} is a tuple $\sys=\system$, where
\begin{itemize}
\item $\nstates$ and $\pstates$ 
are finite disjoint sets of \emph{action} and
\emph{probabilistic states}, resp.,
\item $\atransym\subseteq
\nstates \times \actset \times \nstates \cup \pstates$ is an
\emph{action transition relation} such that $(s,a,t) \in \atransym$
and $(s,a,t') \in \atransym$ implies $t=t'$, and
\item $\ptransym \subseteq
\pstates \times (0,1] \times \nstates$ is a \emph{probabilistic
transition relation} such that, for all $s \in \pstates$,
$\sum_{(s,\pi,t)\in \ptransym}\pi =1$.

\end{itemize}
\end{definition}

\noindent We denote $\nstates \cup \pstates$ by $\states$. We write
$s \atran{a} t$ rather than $(s,a,t) \in \atransym$, and $s
\ptran{\pi} t$  rather than $(s,\pi,t)\in \ptransym$ (or $s \ptran{}
t$ if the value of $\pi$ is irrelevant in the context).  We write $s
\atran{a}$ to denote that there exists an action transition $s
\atran{a}s'$ for some $s' \in \states$. We agree that a state
without outgoing transitions
belongs to $\nstates$.
Given a process $s$ and
action $a \in \actset$, denote by $s_a$ the process, if it exists, for
which $s \atran{a}s_a$. Given a PTS $\sys = \system$, let $I\colon
\nstates \mapsto 2^\actset$ be a function such that, for all $a \in
\actset, s \in \nstates$, it holds $a \in I(s)$ iff $s \atran{a}$.
$I (s)$ is
called the \emph{menu} of $s$. Intuitively, for $s \in \nstates$,
$I(s)$ is the set of actions that the process  $s$ can perform
initially.

As standard, we define a \emph{process graph} (or simply
\emph{process}) to be a state $s \in \states$ together with all
states reachable from $s$, and the transitions between them. A
process graph is usually named by its \emph{root} state, in this
case $s$. 

%% file: ready.tex
\section{Probabilistic ready trace semantics}\label{sec:barbed}

In this section we recall the ready-trace equivalence for reactive
probabilistic processes defined in ~\cite{fossacs10}.

\begin{definition}[Ready trace] A \emph{ready trace of length $n$} is a  sequence
 $(
 \menu_1, a_1, \menu_2, a_2, \ldots, \menu_{n-1}, a_{n-1}, \menu_n)$ where $\menu_i \in 2^\actset$ for all
$i \in \{1, 2, \ldots , n\}$ and $a_i \in \menu_i$ \ for all $i \in
\{1,2, \ldots, n-1\}$ . 
\end{definition}

\noindent We assume that  the observer has the ability to observe
the actions that the process performs, together with the menus out
of which actions are chosen. Intuitively, a ready trace $\obs = (
 \menu_1, a_1, \menu_2, a_2, \ldots, \menu_{n-1}, a_{n-1}, \menu_n)$ can be
observed if the initial menu is $\menu_1$,  then action $a_1 \in
\menu_1$ is performed, then the next menu is $\menu_2$,  then action
$a_2 \in \menu_2$ is performed and so on, until the observing ends
at a point when the  menu is $\menu_n$. It is essential that, since
the probabilistic transitions are not observable, the observer
cannot infer where exactly they happen in the ready trace.

Clearly the probability of observing a ready trace
$(\{a,b\},a,\{c\})$ is conditioned on choosing the action $a$ from
the menu $\{a,b\}$.
 This suggests that, when
defining probabilities on ready traces, the Bayesian definition of
probability is more appropriate than the measure-theoretic
definition that is usually taken.

Next, given a  process $s$, we define a process $s_{(\menu, a)}$.
Intuitively, $s_{(\menu, a)}$ is  the process that $s$ becomes,
assuming that menu $\menu$ was offered to $s$ and action $a$ was
performed.

\begin{definition}\label{def:pafter}
Let $s$ be a process graph. Let $\menu \subseteq \actset$, $a \in
\menu$ be such that $I(s)=\menu$ if $s \in \nstates$ or otherwise
there exists a transition $s \ptran{}s'$ such that $I(s')=\menu$.
The process graph $s_{(\menu,a)}$ is obtained from $s$ in the
following way:
\begin{itemize} \item if $s \in \nstates$ then the root of
$s_{(\menu,a)}$ is the state $s'$ such that $s \atran{a} s'$, and
\item if $s \in \pstates$ then a new state $s_{(\menu, a)}$ is
created. Let $\pi = $
$\sum_{s\ptran{\pi_i}s_i,I(s_i)=\menu}{\pi_i}$. For all $s_i'$ such
that $s \ptran{\pi_i}s_i\atran{a}s_i'$ and $I(s_i)=\menu$:

\begin{itemize} \item if $s_i' \not \ptran{}$, then an edge
$s_{(\menu,a)}\ptran{\pi_i/\pi}s_i'$  is created;
\item for all transitions $s_i' \ptran{\rho_i}s_i''$, an edge
$s_{(\menu,a)}\ptran{\pi_i \rho_i/\pi}s_i''$  is created.
\end{itemize}
\end{itemize}
\end{definition}

\begin{definition} \label{def:P}Let $(
 \menu_1, a_1, \menu_2, a_2, \ldots, \menu_{n-1}, a_{n-1}, \menu_n)$
 be a ready trace of length $n$ and $s$ be a process graph.
 Functions $P_s^1(\menu )$  and
 $\prob_s^n(\menu_n | \menu_1, a_1, \ldots \menu_{n-1}, a_{n-1})$ (for $n > 1$)
 are defined in the following way:\\

\noindent$ \prob_s^1(\menu )=
\begin{cases}{}
\sum_{s \ptran{\pi}s'}{\pi \cdot P_{s'}^1(
\menu )} & \textrm{if } s \in \pstates,\\
1 & \textrm{if } s \in \nstates, \ I(s)=\menu, \\
0 & \textrm{otherwise. } \\
\end{cases}$\\\\
\noindent$\prob_s^2(\menu_2 | \menu_1, a_1) =
\begin{cases}{}
\prob_{s_{(\menu_1,a_1)}}^1(\menu_2) & \textrm{if} \ \prob_s^1(\menu_1)>0, \\
\textrm{undefined} & {otherwise}.
\end{cases}
$\\\\
\noindent$\prob_s^n(\menu_n | \menu_1, a_1, \ldots, \menu_{n-1},
a_{n-1})=
\begin{cases}{} \prob_{s_{(\menu_1,a_1)}}^{n-1}(\menu_n | \menu_2,
a_2, \ldots, \menu_{n-1},
 a_{n-1}) & \textrm{if} \ \prob_s^1(\menu_1)>0, \\
\textrm{undefined} & {otherwise}.
\end{cases}
$
\end{definition}

Let the sample space consist of all possible menus and $s \in
\states$. Function $P_s^1(\menu)$ can be interpreted as the
probability that the menu $\menu$ is observed initially when process
$s$ starts executing.  Let the sample space consist of all ready
traces of length $n$ and let $s \in \states$. The function
$P_s^n(\menu_n | \menu_1, a_1, \ldots \menu_{n-1}, a_{n-1})$ can be
interpreted as the probability of the event $ \{(\menu_1, a_1,
\ldots, \menu_{n-1}, a_{n-1}, \menu_n) \} $, given the event

\noindent $\{(\menu_1, a_1, \ldots \menu_{n-1}, a_{n-1}, \xmenu)
\textrm{~:~} \xmenu \in 2^{\actset} \}$, if observing ready traces
of process $s$. It can be checked that these probabilities are well
defined, i.e., they satisfy the axioms A1-A3 of Section
\ref{sec:prelim}.

\begin{definition}[Probabilistic ready trace
equivalence]\label{def:barbed} Two processes $s$ and $\bar{s}$ are
\emph{probabilistically ready trace equivalent}, notation $s \barbed
\bar{s}$, iff:
\begin{itemize}
\item for all $\menu$ in $2^\actset$, $P_s^1(\menu)=P_{\bar{s}}^1(\menu)$ and
\item for all $n > 1$, $P_s^n(\menu_n | \menu_1, a_1, \ldots \menu_{n-1},
a_{n-1})$ is defined if and only if $P_{\bar{s}}^n(\menu_n |
\menu_1, a_1, \ldots \menu_{n-1}, a_{n-1})$ is defined, and in case
they are both defined,  they are equal.
\end{itemize}
\end{definition}

\noindent Informally, two processes $s$ and $\bar{s}$ are
ready-trace equivalent iff for every $n$ and every ready trace

\noindent $(\menu_1, a_1, \menu_2, a_2 , \ldots \menu_n )$, the
probability to observe $\menu_n$, under condition that previously
the sequence $(\menu_1, a_1, \menu_2, a_2 , \ldots a_{n-1} )$ was
observed, is defined at the same time for both $s$ and $\bar{s}$;
moreover, in case both probabilities are defined, they coincide.
Note that it is straightforward to construct a  testing scenario in
the lines of \cite{spectrum1, CSV07} for this ready-trace
equivalence. Namely, in \cite{spectrum1} a ready trace machine is
described, that allows for the ready traces to be observed. To
\emph{estimate}\footnote{We emphasize the word ``estimate'', as it
is common knowledge that statistics provides only estimations of the
probabilities \cite{lindley}.} the conditional probabilities of the
ready traces of length $n$, only basic statistical analysis needs to
be applied to the set of all ready traces obtained from the
ready-trace machine.

\begin{example} For processes $s$ and $\bar{s}$ in Fig. \ref{fig_machine}
it holds $s \barbed \bar{s}$.
\end{example}

%% file: testing_def.tex
\section{Testing equivalence} \label{sec:testing}

In this section we define a testing equivalence in the style of~\cite{DH84} for reactive probabilistic processes.

Recall  that a division of two polynomials is called a
\emph{rational function}. For example, $\frac{2x}{x+y}$ is a
rational function with  arguments $x$ and $y$. A possible domain for
this function is $(0, \infty) \times (0, \infty)$. We exploit a
subset $\ratexpr$ of the rational functions
 whose argument names
belong to the action labels $\actset$, which is generated by the
following grammar:
\[\varphi::=\alpha  \ \mid \   a \ \mid \ \varphi + \varphi \ \mid \
\varphi \cdot \varphi \ \mid \ \frac{\varphi}{\varphi},\] \noindent
where $\alpha$ is a non-negative scalar, $a \in \actset$, and $+, \
\cdot, \ $ and $ \frac{\cdot}{\cdot}$ are ordinary algebraic
addition, multiplication and fraction, resp. Brackets are used in
the standard way to change the priority of the operators. For our
purposes, we assume that the arguments  $a,b, ...$ can only take
positive values, i.e. the domain of every function in $\ratexpr$ is
$(0, \infty)^{n}$, where $n$ is the size of the action set.
Therefore, two rational functions in $\ratexpr$ are equal iff they
can be transformed to equal terms using the standard transformations
that preserve equivalence (e.g. for $a,b \in \actset$, $
\frac{1}{2}\cdot \frac{a}{a+b}+\frac{1}{2}\cdot
\frac{b}{a+b}=\frac{1 \cdot (a+b)}{2 \cdot(a+b)}=\frac{1}{2}$).

A \emph{test} $T$, as standard,  is a finite process~\footnote{For
now we restrict to non-recursive tests, as the characterization
proof in Sec.~\ref{sec:relationship} is already involved; however,
it is not uncommon to restrict to non-recursive
 tests in probabilistic testing initially,
  for the sake of clear presentation (see e.g.~\cite{DGHM08,DGHM09}).
In fact, usually recursive tests do not increase the distinguishing
power of the finite tests
\cite{Hen88,segala_testing96,verena_testing}, since infinite paths
in tests cannot report success.
  } 
 such that,
 for a symbol $\success \not \in \actset$, there may exist
transitions $s \atran{\success} $
 for some  states $s$ of $T$, denoting success. Denote the set of all tests by $\tests$. Next, we define the result of testing a
process with a given test. The informal explanation follows
afterwards.

\begin{definition}\label{def:Results} The function
$\Result \colon \states \times \tests \mapsto \ratexpr$ that gives
the result of testing a process $s$ with a test $T$ is defined as
follows:

\[ \Result(s,T)=
\begin{cases} {}
1, & \text{if } T \atran{\success}, \\

\sum_{i \in I}{\pi_i \cdot \Result(s_i , T)}, &  \text{if } s
\pitran{\pi_i}s_i \text{ for } i \in I$, $T  \not \atran{\success} \\

\sum_{i \in I}{\pi_i \cdot \Result(s , T_i)}, &  \text{if } T
\pitran{\pi_i}T_i \text{ for } i \in I$, $s  \not \ptran{} \\


\sum_{a \in K} \frac{a}{\sum_{b \in K}{b}}  \Result(s_a , T_a), &
\text{for } K \text{=} I(s) \cap I(T),\text{ otherwise}.

\end{cases}
\]
\end{definition}

\noindent As usual, the result of testing a process  with a success
test is $1$. The result of testing a process with a probabilistic
state as a root (i.e. initially probabilistic process) is a weighted
sum of the results of testing the subsequent processes with the same
test. Similarly when the test is initially probabilistic.
Non-standard, however, is in the result of testing
a process $s$ with a test $T$ that can initially
perform actions from $\actset$ only. Namely, when the process and
the test synchronize on an action, the resulting transition is
labeled with a ``weighting factor'', containing information about
the way this synchronization happened.
This information has form of a rational function, the numerator of
which represents the synchronized action itself, while the
denominator is the sum of the common initial actions of $s$ and $T$,
i.e., all actions on which $s$ and $T$ could have synchronized at
the current step. In order to compute the final result of the
testing,  the rational function is temporarily treated as
``symbolic'' probability. The final result is again a rational
function in $\ratexpr$.

For example, it is easy to compute that the result of testing either
$s$ or $\bar{s}$ with $u$ (given in Fig.~\ref{fig_machine}) is equal
to $\frac{1}{2}$, which establishes one of our goals set in
Sec.~\ref{sec:introduction}. However, in many cases the result is a
non-scalar rational function. 

\begin{definition} \label{def:testing} Two processes $s$ and $\bar{s}$ are \emph{testing
equivalent}, notation $s \testing \bar{s}$, iff $\Result(s,T)$ and
$\Result(\bar{s},T)$ are equal functions for every test $T$.
\end{definition}

\noindent Obviously, comparing two results boils down to comparing
two polynomials, after both rational functions have been transformed
to equal denominators.

\begin{remark} In~\cite{fossacs10}, in order to keep the probabilities
in a composed system, the actions resulting from synchronization
have a label containing information about the present and the
history of synchronization -- i.e. a sequence of previous menus of
actions-candidates for synchronizing and the actual synchronized
actions. This is because (i) we would like to denote that both
choices in $s || u$ (Fig.~\ref{fig_machine}) are resolved in the
same way and (ii) the history of resolution of choices, as usual,
can play a role in the current resolution.
In the present paper one of our main goals is to prove that the
testing equivalence coincides with the ready-trace equivalence for
the model of reactive probabilistic processes.\footnote{In the
setting without internal nondeterminism, preorder relations become
superfluous, since in \cite{fossacs10}, as usual, a process
\emph{implements} another one iff the former contains less internal
nondeterminism} It turns out that, in order to achieve this goal, we
can simplify the notation for the label of a synchronized action.
Here the label of a resulting synchronized action contains only information about the
current circumstances of synchronization in the form of a rational
expression and the result of testing remains a rational expression.
(The rational function is a suitable form of ``remembering'' the
information, because ``in the world of rational expressions''
commutativity and distributivity laws hold, analogous to those we
try to achieve ``in the world of processes''.) Besides simplifying
the notation, this labeling enables us to present the proof of
Theorem~\ref{thm:levo_desno} (Sec.~\ref{sec:relationship}) in a much
more concise way.
\end{remark}

%% file: relationship.tex
\section{Relationship between $\testing$ and
$\barbed$}\label{sec:relationship}

We establish our main result, namely that the testing equivalence
$\testing$ coincides with the probabilistic ready-trace equivalence
$\barbed$. In \cite{fossacs10},  given that two processes are not
ready-trace equivalent, we provide a procedure on how to construct a
test that distinguishes between the processes. The procedure heavily
relies on the fact that tests can perform internal transitions
(which can be manipulated based on the synchronization history). In
the present case, the internal transitions of the tests, as those of
the processes, are fully random (both the tests and the processes
belong to the same model, in the spirit of \cite{DH84}). Because of
this, the present characterization proof is rather based on
contradiction and is much more technically involved.

As an intermediate result, we prove that the probabilistic
transitions do not add distinguishing power to the tests.


The following lemma, which considers the determinant of a certain
type of an almost-triangular matrix, shall be needed in the proof of
Theorem \ref{thm:levo_desno}.

\begin{lemma} \label{lem:determinant}Let $\mathbf{Q}$ be a square $n \times n$ matrix
with elements $q_{ij}$, for $1\leq i \leq n$ and $1\leq j \leq n$.
Suppose $q_{ij}\in \{0,1\}$ for $i>1$, $q_{ij}=1$ \ for $i=j+1$,
$q_{ij}=0$ for $i>j+1$, and $q_{1j}=\frac{Q_1}{Q_j}$ for $1\leq j
\leq n$, where $Q_1, Q_2\ldots Q_n$ are  irreducible, mutually prime
polynomials with positive variables, and of non-zero degrees. Then
the determinant of $\mathbf{Q}$ is a non-zero rational function.
\end{lemma}

\begin{proof}
The determinant $\Det(\mathbf{Q})$ of  matrix $\mathbf{Q}$ can be
obtained from the general recursive formula
$\Det(\mathbf{Q})=\sum_{j=1}^{n}(-1)^{1+j}q_{1j}\Det(\mathbf{Q_{1j}})
$, where $\mathbf{Q_{1j}}$  is  the matrix obtained by deleting the
first row and the $j$-th column of $\mathbf{Q}$. Observe that
$\mathbf{Q_{1n}}$ is an upper-triangular matrix, the diagonal
elements of which are all equal to one. Since the determinant of a
triangular matrix is equal to the product of its diagonal elements,
we have $\Det(\mathbf{Q_{1n}})=1$. Therefore, the coefficient in
front of the rational function $\frac{Q_1}{Q_n}$ in
$\Det(\mathbf{Q})$ is equal to $1$. Suppose $\Det(\mathbf{Q})$ is a
zero-function. Then, the rational function $\frac{1}{Q_n}$ is equal
to a
 linear combination of $\frac{1}{Q_1}, \ldots \frac{1}{Q_{n-1}}$. This means  that the rational function $\frac{Q_1 \cdot
Q_2\cdot \ldots \cdot Q_{n-1}}{Q_n}$ is a polynomial. The last
 is impossible,
since, by assumption, the denominator is an irreducible polynomial
of non-zero degree and is not contained in the numerator. Therefore,
$\Det(\mathbf{Q})$ is not a zero-function.
\end{proof}

\begin{theorem} \label{thm:levo_desno}
 Let $s$ and $t$ be two processes such that $s \not \barbed
t$. There exists a test $T$ that has no probabilistic transitions
such that $\Result(s,T) \not = \Result(t,T)$.
\end{theorem}

\begin{proof}
We prove the theorem by induction on the minimal length $m$ of a
ready-trace that distinguishes between $s$ and $t$. For $m=1$, we
prove that the test $T=\extch_{a\not \in \menu}{a.\success}$, where
$\menu$ is a menu with a minimal possible number of actions such
that $\prob_s^1(\menu)\not = \prob_t^1(\menu)$, distinguishes
between $s$ and $t$. For $m>1$ the proof goes as follows. If
$\prob_s^1(\menu) = \prob_t^1(\menu)$ for every menu $\menu$, then
by the inductive assumption it follows that there exists a test
$T_1$, menu $\menu_1$ and action $a_1 \in \menu_1$ such that
$\Result(s_{(\menu_1,a_1)},T_1)
 \not  = \Result(t_{(\menu_1,a_1)},T_1)$. We show that there exists
 a subset of the action set, say $\mathsf{Act}$, such that the test $T=a_1.T_1+\extch_{b \in
 \mathsf{Act}}b.\success$ distinguishes between $s$ and $t$. To prove this,
we take $\menu_1$ to be
 the menu containing a minimal possible number of actions such that $\prob_s^1(\menu_1)>0$, $a_1 \in \menu_1$,  and
$\Result(s_{(\menu_1,a_1)},T_1)
 \not  = \Result(t_{(\menu_1,a_1)},T_1)$. Then we take the set
 $\mathsf{Act'}$ to consist of the actions that can be initially
 performed by $s$ but do not belong to menu $\menu_1$. Then, we show
 that there must exist a subset $\mathsf{Act}$ of $\mathsf{Act'}$
 such that the test $T=a_1.T_1+\extch_{b \in
 \mathsf{Act}}b.\success$ distinguishes between $s$ and $t$ (otherwise, we obtain that $\Result(s_{(\menu_1,a_1)},T_1)
   = \Result(t_{(\menu_1,a_1)},T_1)$, which contradicts our assumption).

We now proceed with a detailed presentation of the proof.

 From $s \not \barbed t$ and by Def. \ref{def:barbed}, there
must exist a ready-trace $(\menu_1, a_1, \ldots \menu_m)$ such that

\noindent $\prob_s^m(\menu_m | \menu_1, a_1, \ldots a_{m-1}) \not =
\prob_t^m(\menu_m | \menu_1, a_1, \ldots a_{m-1})$. The proof is by
induction on $m$.\\

\textbf{Case 1 ($m=1$)} Suppose first that there exists a menu
$\menu$ such that
$\prob_s^1(\menu)\not = \prob_t^1(\menu)$. 
Let $\menu$ be a menu with a minimal possible number of actions such
that $\prob_s^1(\menu)\not = \prob_t^1(\menu)$. Take
$T=\extch_{a\not \in \menu}{a.\success}$. We have $\Result(s,T) = 1-
\sum_{\menu' \subseteq \menu}\prob_s^1(\menu')$, because the actions
of $s$ and $T$ will fail to synchronize if and only if the random
choice decides that menu $\menu$ or some menu $\menu' \subset \menu$
is offered to process $s$ initially. Similarly, $\Result(t,T) = 1-
\sum_{\menu' \subseteq \menu}\prob_t^1(\menu')$. Now, suppose that
$\Result(s,T)
  = \Result(t,T)$. We have
   $\sum_{\menu' \subseteq \menu}\prob_s^1(\menu')=
   \sum_{\menu' \subseteq \menu}\prob_t^1(\menu')$. From this and  $\prob_s^1(\menu)\not =
\prob_t^1(\menu)$, it follows that there  exists a menu $\menu'
\subset \menu$ such that also $\prob_s^1(\menu')\not =
\prob_t^1(\menu')$. But this contradicts the assumption that $\menu$
is a menu with a minimal possible number of actions such that
$\prob_s^1(\menu)\not =
\prob_t^1(\menu)$.\\

\textbf{Case 2 ($m>1$)} Suppose now that $\prob_s^1(\menu) =
\prob_t^1(\menu)$ for every menu $\menu$. Let $(\menu_1, a_1, \ldots
\menu_m)$ be a ready-trace
such that 
$\prob_s^{m}(\menu_m | \menu_1, a_1, \ldots a_{m-1}) \not =
\prob_t^{m}(\menu_m | \menu_1, a_1, \ldots a_{m-1})$. From
$\prob_s^1(\menu_1) = \prob_t^1(\menu_1)$, and from  Definitions
\ref{def:pafter} and \ref{def:P}, it follows that
$\prob_{s_{(\menu_1,a_1)}}^{m-1}(\menu_m | \menu_2, a_2, \ldots
a_{m-1}) \not = \prob_{t_{(\menu_1,a_1)}}^{m-1}(\menu_m | \menu_2,
a_2, \ldots a_{m-1})$ (in case $m=2$,
$\prob_{s_{(\menu_1,a_1)}}^1(\menu_2 ) \not =
\prob_{t_{(\menu_1,a_1)}}^1(\menu_2) $). Now, by the inductive
assumption, there exists a non-probabilistic test $T_1$ such that
$\Result(s_{(\menu_1,a_1)},T_1)
 \not  = \Result(t_{(\menu_1,a_1)},T_1)$.\\

\textbf{Case 2.1} \ Suppose first that $a_1$ does not belong to any
first-level menu of $s$ other than $\menu_1$, i.e. that for every
menu $\menu$, $\prob_s^1(\menu)>0$ and $a_1 \in \menu$ implies
$\menu = \menu_1$.  Then the test $T=a_1.T_1$ distinguishes
between $s$  and $t$.\\

\textbf{Case 2.2} \ Suppose now that $a_1$ belongs to at least one
first-level menu of
 $s$ other than $\menu_1$, i.e.
there exists at least one menu $\menu \not = \menu_1$ such that
$\prob_s^1(\menu)>0$ and $a_1 \in \menu$.
 Without loss of
 generality, assume that $\menu_1$ is a menu with a minimal possible number of actions such that
$\prob_s^1(\menu_1)>0$, $a_1 \in \menu_1$,  and
$\Result(s_{(\menu_1,a_1)},T_1)
 \not  = \Result(t_{(\menu_1,a_1)},T_1)$.
 Let $\{b_j\}_{j \in J}$ be
the set of actions that appear in the first level of $s$ (and
therefore $t$) but not in $\menu_1$, i.e. $b \in \{b_j\}_{j \in J}$
if and only if $b \not \in \menu_1$ and there exists a menu $\menu$
such that $\prob_s^1(\menu)>0$, $b \in \menu$. We shall prove that
there exists $J' \subseteq J$ such  that the test $T=a_1.T_1+
\extch_{j \in J' }b_j.\success$ distinguishes between $s$ and $t$.
More concretely, we shall prove that, assuming the opposite, it
follows that $\Result(s_{(\menu_1,a_1)},T_1)
  = \Result(t_{(\menu_1,a_1)},T_1)$, thus obtaining contradiction.
  \\

\textbf{Case 2.2.a} \ Suppose first that $\{b_j\}_{j \in J} =
\emptyset$. This means that there are no actions other than those in
$\menu_1$, that appear in the first level of $s$. Therefore, all
menus $\menu$ for which $\prob_s^1(\menu)>0$ satisfy $\menu
\subseteq \menu_1$. We prove that the test $T=a_1.T_1$ distinguishes
between $s$ and $t$. Assume that $\Result(s,T)=\Result(t,T)$. From
the last and from Def. \ref{def:Results}, we obtain
\begin{align}\label{eq:submenus}
\sum_{\menu:\prob_s^1(\menu)>0,a_1\in\menu \subseteq
\menu_1}(\Result (s_{(\menu,a_1)},T_1)-\Result
(t_{(\menu,a_1)},T_1)) =0.
\end{align}
 By
assumption, for every $\menu \subset \menu_1$ such that $a_1 \in
\menu$ it holds $\Result (s_{(\menu,a_1)},T_1)=\Result
(t_{(\menu,a_1)},T_1)$. Therefore, from \eqref{eq:submenus} we
obtain $\Result (s_{(\menu_1,a_1)},T_1)=\Result
(t_{(\menu_1,a_1)},T_1)$, which contradicts the assumption $\Result
(s_{(\menu_1,a_1)},T_1)\not =\Result (t_{(\menu_1,a_1)},T_1)$.\\

\textbf{Case 2.2.b} \ Suppose now that $\{b_j\}_{j \in J} \not =
\emptyset$.
 Given action
$b_i \in \{b_j\}_{j \in J}$, denote by $\mathcal{M}_i$ the set of
all first-level menus of $s$ that contain $b_i$ and $a_1$, i.e.
$\menu \in \mathcal{M}_i$ iff $\prob_s^1(\menu) >0$ and $b_i,a_1 \in
\menu$; denote by $\mathcal{M}_i^C$ the set of all first-level menus
of $s$ that do not contain $b_i$ but have $a_1$, i.e. $\menu \in
\mathcal{M}_i^C$ iff $\prob_s^1(\menu) >0$, $b_i \not \in \menu$ and
$a_1 \in \menu$.

Let $T=a_1.T_1+\extch_{j \in J' }b_j.\success$ for some
$J'=\{1,2,\ldots n\} \subseteq J$ and suppose $\Result(s,T)
  = \Result(t,T)$. Since $\prob_s^1(\menu)=\prob_t^1(\menu)$ for every menu $\menu$,
  observe that only if action $a_1$ is performed initially, it is
possible for the test $T=a_1.T_1+\extch_{j \in J' }b_j.\success$  to
make a difference between $s$ and $t$. Because of this and by
Definitions \ref{def:Results} and \ref{def:pafter}  it follows  that
 \begin{align}\label{eq:thebig}
\sum_{\menu \in \mathcal{M}_n^C \cap \mathcal{M}_{n-1}^C \cap \cdots
\cap
\mathcal{M}_1^C}\frac{a_1}{a_1}\prob_s^1(\menu)\varpi(\menu) 
+\sum_{\menu \in \mathcal{M}_n^C \cap \mathcal{M}_{n-1}^C \cap
\cdots \cap
\mathcal{M}_1}\frac{a_1}{a_1+b_1}\prob_s^1(\menu)\varpi(\menu)
+\cdots \notag\\
+\sum_{\menu \in \mathcal{M}_n   \cap \cdots \cap
\mathcal{M}_1}\frac{a_1}{a_1+\sum_{j=1}^{n}b_j}\prob_s^1(\menu)\varpi(\menu)
=  0,
\end{align}
where by $\varpi(\menu)$ we denote
$\Result(s_{(\menu,a_1)},T_1)-\Result(t_{(\menu,a_1)},T_1).$
 \noindent Each intersection appearing
under the $\sum$-operators of \eqref{eq:thebig} can be mapped
bijectively to a binary number of $n$ digits -- the i-th digit being
$0$ if the intersection contains $\mathcal{M}_{n+1-i}^C$, and $1$ if
the intersection contains $\mathcal{M}_{n+1-i}$. (For reasons that
will become clear later, the order of the indexing is reversed.)

Suppose  $\Result(s,T) = \Result(t,T)$ for every test
$T=a_1.T_1+\extch_{j \in J' }b_j.\success$, where $J'\subseteq J$.
We shall prove that, in this case, every sum $\sum\varpi(M)$ that
appears in \eqref{eq:thebig} when $J'=J$ is equal to a
zero-function. In particular, the equality
\begin{equation}\label{eq:presek}
\sum_{\menu \in \bigcap_{j \in J}\mathcal{M}_j^C}\varpi(M)=0
\end{equation}
\noindent will hold. Note  that the set $\bigcap_{j \in
J}{\mathcal{M}_j^C}$ contains all first-level menus of $s$ that have
the action $a_1$ but do not have any other action that does not
appear in $\menu_1$. Therefore, $\bigcap_{j \in J}{\mathcal{M}_j^C}$
consists of the subsets of $\menu_1$ that contain $a_1$. Thus, the
equation \eqref{eq:presek} is equivalent to the equation
\eqref{eq:submenus} which leads to $\Result(s_{(\menu_1,a_1)},T_1)
=\Result(t_{(\menu_1,a_1)},T_1)$, i.e. to  contradiction. This would complete the proof of the theorem.\\

We now proceed with proving the above stated claim. We prove a more
general result,  namely that for $J' \subseteq J$, under assumption
that $\Result(s,T) = \Result(t,T)$ for every test
$T=a_1.T_1+\extch_{i \in J'' }b_i.\success$ such that $J'' \subseteq
J$ and $|J''|\leq|J'|$, it holds that every sum $\sum\varpi(M)$
that appears in \eqref{eq:thebig} is equal to zero. \\

Suppose first that $|J'|=1$, i.e. $J'=\{1\}$. Assume that
\begin{equation}\label{eq:eden1}
\Result(s,a_1.T_1) = \Result(t,a_1.T_1)
\end{equation}
  and
\begin{equation}\label{eq:eden2}
\Result(s,a_1.T_1+b_1.\success) = \Result(t,a_1.T_1+b_1.\success).
\end{equation}
 From \eqref{eq:eden1}, Def. \ref{def:Results},
  and because $\prob_s^1(\menu)=\prob_t^1(\menu)$ for every menu $\menu$, we  obtain
\begin{align}\label{eq:b1}
\sum_{\menu \in \mathcal{M}_1 \cup \mathcal{M}_1^C
}\frac{a_1}{a_1}\prob_s^1(\menu)(\Result(s_{(\menu,a_1)},T_1)-\Result(t_{(\menu,a_1)},T_1))
 =0.
\end{align}
The equation \eqref{eq:thebig} turns into
\begin{align}\label{eq:b2}
& \sum_{\menu \in
\mathcal{M}_1^C}\frac{a_1}{a_1}\prob_s^1(\menu)\varpi(M)
\textrm{+}\sum_{\menu \in
\mathcal{M}_1}\frac{a_1}{a_1+b_1}\prob_s^1(\menu)\varpi(M) =  0.
\end{align}
Denote $\sum_{\menu \in \mathcal{M}_1^C}\prob_a^1(\menu)\varpi(M)$
by $x_0$ and $\sum_{\menu \in
\mathcal{M}_1}\prob_a^1(\menu)\varpi(M)$ by $x_1$. Our goal is to
show  that $x_0=0$ and $x_1=0$, i.e. that they are zero-functions.
From \eqref{eq:b1} and \eqref{eq:b2} we obtain the system of
equations for the unknowns $x_0$ and $x_1$
%
%
  $$ \mathbf{Q_1}\mathbf{x}=\mathbf{0},$$ where
$$\mathbf{Q_1}=\left(
                      \begin{array}{cc}
                        \frac{a_1}{a_1} & \frac{a_1}{a_1+b_1} \\
                        1 & 1 \\
                      \end{array}
                    \right), \mathbf{x}=\left(
                                            \begin{array}{c}
                                              x_0 \\
                                              x_1 \\
                                            \end{array}
                                          \right)
                    , \textrm{ and } \mathbf{0}=\left(
                                         \begin{array}{c}
                                           0 \\
                                           0 \\
                                         \end{array}
                                       \right).
                    $$
Since the determinant of the matrix $\mathbf{Q_1}$ is not a
zero-function, it follows that $x_0=0$ and $x_1=0$ is the only
solution of the system.

To present a better intuition on the proof in the general case, we
shall also consider separately  the case $|J'|=2$. Let $J'=\{1,2\}$
and assume that $\Result(s,T) = \Result(t,T)$ for every test
$T=a_1.T_1+\extch_{i \in J'' }b_i.\success$ such that $J'' \subseteq
J$ and $|J''|\leq|J'|$. The equation \eqref{eq:thebig} turns into

 \begin{align}\label{eq:thebig2}
& \sum_{\menu \in \mathcal{M}_2^C \cap
\mathcal{M}_1^C}\frac{a_1}{a_1}\prob_s^1(\menu)\varpi(\menu)
+ \sum_{\menu \in \mathcal{M}_2^C \cap \mathcal{M}_1}\frac{a_1}{a_1+b_1}\prob_s^1(\menu)\varpi(\menu)  \notag\\
+& \sum_{\menu \in \mathcal{M}_2 \cap
\mathcal{M}_1^C}\frac{a_1}{a_1+b_2}\prob_s^1(\menu)\varpi(\menu) +
\sum_{\menu \in \mathcal{M}_2 \cap
\mathcal{M}_1}\frac{a_1}{a_1+b_1+b_2}\prob_s^1(\menu)\varpi(\menu)
=0.
\end{align}
Denoting   $\sum_{\menu \in \mathcal{M}_2^C \cap \mathcal{M}_1^C
}\prob_s^1(\menu)\varpi(M)$ by $x_{00}$ and so on,
\eqref{eq:thebig2} turns into
\begin{align}
\frac{a_1}{a_1}x_{00} + \frac{a_1}{a_1+b_1}x_{01} +
\frac{a_1}{a_1+b_2} x_{10} + \frac{a_1}{a_1+b_1+b_2} x_{11}=0.
\end{align}
From $\sum_{\menu \in \mathcal{M}_2^C}\prob_s^1(\menu)\varpi(M)=0$
we obtain $x_{00}+x_{01}=0$, and from $\sum_{\menu \in
\mathcal{M}_2}\prob_s^1(\menu)\varpi(M)=0$ we obtain
$x_{10}+x_{11}=0$. Similarly, from $\sum_{\menu \in
\mathcal{M}_1}\prob_s^1(\menu)\varpi(M)=0$ we obtain that
$x_{01}+x_{11}=0$. Therefore, we have the system
$\mathbf{Q_2x}=\mathbf{0},$ where
$$\mathbf{Q_2}= \left(
  \begin{array}{cccc}
    \frac{a_1}{a_1} & \frac{a_1}{a_1+b_1} & \frac{a_1}{a_1+b_2} & \frac{a_1}{a_1+b_1+b_2} \\
    1 & 1 & 0 & 0 \\
    0 & 1 & 0 & 1 \\
    0 & 0 & 1 & 1 \\
  \end{array}
\right). $$ By Lemma \ref{lem:determinant}, $\Det(\mathbf{Q_2})$ is
not a zero-function, which implies that the vector of zero-functions
is
the only solution of the above system of equations.\\

We now present how each matrix $\mathbf{Q_{n+1}}$ can be obtained
from the matrix $\mathbf{Q_{n}}$.

In general, for $\mathcal{M}_i^* \in
\{\mathcal{M}_i,\mathcal{M}_i^C\}$, it holds
\begin{align}
\sum_{\menu \in (\bigcap_{i=1}^n{\mathcal{M}_i^*}) \cap
\mathcal{M}_{n+1}} \prob_s^1(\menu)\varpi(\menu) + \sum_{\menu \in
(\bigcap_{i=1}^n{\mathcal{M}_i^*}) \cap \mathcal{M}_{n+1}^{C}}
\prob_s^1(\menu)\varpi(\menu) =  \sum_{\menu \in
(\bigcap_{i=1}^n{\mathcal{M}_i^*})} \prob_s^1(\menu)\varpi(\menu).
\end{align}
This means that, in the general case, each solution $x_{i_1 i_2
\ldots i_n}=0$ of the system $\mathbf{Q_n}\mathbf{x}=\mathbf{0}$
generates  the following equations for the next system: \[x_{i_1 i_2
\ldots i_{k} 0 i_{k+1} \ldots i_{n}} + x_{i_1 i_2 \ldots i_{k} 1
i_{k+1} \ldots
i_{n}}=0,\] for every $0\leq k \leq n$. 
Note that each row of $\mathbf{Q_2}$, except the first one, contains
exactly two $1$'s, at positions whose binary representations differ
in exactly one place.

Informally, the general algorithm for obtaining the elements
$q_{n+1}^{ij}$ of a $2^{n+1}\times 2^{n+1}$ matrix
$\mathbf{Q}_{n+1}$ from matrix $\mathbf{Q}_n$, assuming
$\mathbf{Q}_n$ is non-singular, is as follows. First, initialize all
elements of $\mathbf{Q}_{n+1}$ to zero. Then, copy $\mathbf{Q}_{n}$
into the upper left corner of $\mathbf{Q}_{n+1}$. Then, copy
$\mathbf{Q}_{n}$, excluding the first row, into the lower right
corner of $\mathbf{Q}_{n+1}$. Then, assign $1$ to $q_{n+1}^{ij}$ for
$i=2^n+1$ and $j \in \{2^n, 2^{n+1}\}$. Finally, add the appropriate
new rational fractions in the second half of the first row of
$\mathbf{Q}_{n+1}$. The key observation is that in this way, we
obtain again a matrix such that each row, except the first one,
contains exactly two $1$'s, at positions whose binary
representations differ in exactly one place. Formally,
\[ q_{n+1}^{ij}=
\begin{cases}{}
q_n^{ij} & \textrm{if } 1 \leq i \leq 2^n \textrm{ and } j \leq 2^n, \\
1 & \textrm{if } i=2^n+1 \textrm{ and } j \in \{2^n, 2^{n+1}\}, \\
q_n^{ij} & \textrm{if } 2^n+1<i \textrm{ and } 2^n <j, \\
\frac{a_1}{a_1+\sum_{k \in K}{b_k}+b_{n+1}} & \textrm{if } i=1,j>2^n, \\ & \textrm{ and }  q_n^{(i) (j-2^n)}= \frac{a_1}{a_1+\sum_{k \in K}{b_k}}\\
0 & \textrm{otherwise}.
\end{cases}
\]
Assuming matrix $\mathbf{Q}_{n}$ satisfies the conditions of Lemma
\ref{lem:determinant}, it easily follows that  matrix
$\mathbf{Q}_{n+1}$ also satisfies the conditions of Lemma
\ref{lem:determinant}. Therefore, its determinant is  not a zero
function. This means that the system
$\mathbf{Q}_{n+1}\mathbf{x}=\mathbf{0}$ has only zero-functions as
solutions, which we were aiming to prove. Therefore, the proof of
the theorem is complete.
\end{proof}

\begin{theorem}\label{thm:desno_levo}
Let $s$ and $t$ be two processes. If $s \barbed t$ then $s \testing
t$.
\end{theorem}
\begin{proof} Straightforward: see \cite{qapl10_report}.
\end{proof}

From Theorems \ref{thm:desno_levo} and \ref{thm:levo_desno} the
following statements directly follow.
\begin{theorem} For arbitrary processes $s$ and $t$,  $s \testing t$
if and only if $s \barbed t$.
\end{theorem}

\begin{theorem}\label{thm:nonprobabilistic} For arbitrary processes $s$ and $t$,  $s \not \testing
t$ if and only if there exists a test $T$ without probabilistic
transitions such that $\Result(s,T)\not = \Result(t,T)$.
\end{theorem}

\begin{remark}
It is interesting to note that, while in the non-probabilistic case
the may/must  testing equivalence can be characterized with the
failure equivalence \cite{D87}, in the probabilistic case we obtain
a bit finer characterization. However, this is not unusual in the
probabilistic case, due to the ``effect'' of the probabilities --
e.g. the same phenomenon appears also in the fully probabilistic
case \cite{nunez03}.
\end{remark}

%% file: decidability.tex
\section{Testing systems and decidability}\label{sec:decidability}

In this section we outline how testing can be applied to systems for
which only partial information may be known, and we show that the
testing equivalence is decidable for finite systems or up to a
certain depth of the systems.

So far we have discussed testing ``processes'', i.e. models of
systems. In practice, to test a system with a given test, the
probabilistic transitions of the system need not be known. Namely,
assume that when the system and the test are ready to synchronize on
an action, the test can ``see'' the actions-candidates for
synchronization. If the system is tested with the test exhausting
all possible ways of synchronization and sufficiently many times,
then the result shall be a set of rational functions without
scalars; a standard statistical analysis will give an estimation of
the probability distribution over the rational functions. (A
detailed description of the procedure is beyond the goals of the
current paper.) Two systems would not be distinguished under a given
test iff the resulting distributions are the same.
 The assumption
that the test can see the actions-candidates for synchronization, on
the other hand, corresponds to the user (e.g. $u$ in Fig.
\ref{fig_machine}) being able to see the menu that the machine (e.g.
$s$ in Fig. \ref{fig_machine}) offers. Indeed, this assumption does
not exist in the standard non-probabilistic testing
theory~\cite{DH84}. However, in real-life systems this is not
unusual. Moreover,  this assumption is mild with respect to the
probabilistic may/must testing approaches discussed in Sec.
\ref{sec:introduction}, where one needs to have insight into the
internal structure of the composed system in order to determine the
possible schedulers.

From Theorem \ref{thm:nonprobabilistic} it follows that
non-probabilistic, i.e. deterministic tests suffice to distinguish
between two processes. Therefore, since the action set is finite, an
algorithm for deciding equivalence on finite processes, or up to a
certain length, can be easily constructed. Namely, in this case the
characteristic set of tests of a given length is finite. In case the
length of the processes is unknown, the procedure  stops when, for a
certain length of the tests, the testing yields result $0$ for every
test of that length and every tested process (meaning that the
maximal length of the processes has been exceeded).

\begin{proposition} There exists an algorithm that decides
$\testing$ for finite processes.
\end{proposition}

%% file: conclusion.tex
\section{Related work and conclusion}\label{sec:conclusion}

There is a plethora of equivalences defined on probabilistic
processes in the last two decades (e.g.\cite{chr90,LS91,
YL92,segalaPhD,morgan96refinementoriented,cdsy99,DLZ06, CSV07,
Henzinger08}). However, we think that closely related to ours are
the research reports that face the challenge of allowing
unobservability of the internal probabilistic choice, but still not
allowing more identifications than the standard
must-testing~\cite{DH84,Hen88}, if probabilistic choice is treated
as a kind of internal choice~\footnote{See~\cite{spectrum1} for the
properties that the must-testing equivalence preserves, but are not
preserved by a (completed) trace equivalence.}.

Testing equivalences in the style of~\cite{DH84} for processes with
external choice and internal probabilistic choice, that allow
unobservability of the probabilistic choice, i.e. distribution of
prefix over probabilistic choice, have been also defined
in~\cite{morgan96refinementoriented,CCVPP03,CP07}. Of these,
only~\cite{CP07}, under certain conditions, equate  processes $s$
and $\bar{s}$ of Fig.~\ref{fig_machine}. In~\cite{CP07} process
states are enriched with labels, and a testing equivalence is
defined by means of schedulers that synchronize with processes on
the labels. While in our work  processes $s$ and $\bar{s}$ in Fig.
\ref{fig_machine} are equivalent, in ~\cite{CP07} these two
processes can be equated iff the labeling is right.

Probabilistic equivalences in  ready-trace style have been defined
in \cite{Lowe93} and \cite{spancite97}, also for processes where the
internal nondeterminism has been quantified with probabilities.
However, in contrast to our approach, these definitions do not imply
testing scenarios that can characterize the equivalences, as the one
given in \cite{spectrum1} for the non-probabilistic ready-trace
equivalence.

Other equivalences, that also allow distributivity of prefix over
probabilistic choice, but are not closely related to ours, include
trace-style equivalences (~\cite{segalaPhD,Seidel95, CLSV06,CSV07,
Henzinger08}) and button-pushing testing equivalences
(~\cite{KN98b},\cite{LNR06}). Of these, only~\cite{Seidel95},
\cite{Henzinger08} and~\cite{LNR06} also allow distribution of
external choice over probabilistic choice. However, in these
approaches the environment is not a process itself, but rather a
sequence of actions. In other words, their motivation does not
include sensitivity to deadlock and branching structure -- e.g. they
also identify processes $c.a \oplus_{\frac{1}{2}} c.b$ and $c.(a+b)
\oplus_{\frac{1}{2}} c$ (``+'' being the operator for external
choice).

The present paper is also related to the newer research in
\cite{CP07,CLSV06, GD09}, in the sense that it restricts the power
of the schedulers that resolve the nondeterminism in  a parallel
composition. Contrary to   \cite{CP07,CLSV06, GD09}, the
``schedulers'' in the present paper do not use information about the
state in which a process is. We believe that this approach is more
appropriate when defining a testing equivalence on processes, as it
is closer in nature to the work in \cite{DH84}.

Finally, so far, none of the proposals of testing equivalences in
the style of \cite{DH84} for probabilistic processes having
``external nondeterminism''  deal with the problem of deciding
equivalence based on the testing semantics itself. We refer the
reader also to \cite{verena_testing} for a survey of the testing
equivalences on probabilistic processes and decidability results.

To conclude, we have proposed a testing equivalence in the style
of~\cite{DH84} for processes where the internal nondeterminism is
quantified with probabilities (e.g.~\cite{LS91,LNR06}). We showed
that it can be characterized as a probabilistic ready-trace
equivalence. From the characterization it follows that: (i) the
testing equivalence is insensitive to the exact moment of occurrence
of an internal probabilistic choice, (ii) it equates no more
processes than the equivalence of \cite{DH84} when probabilities are
not treated, and (iii) a decidability procedure exists for
determining if two finite processes are testing equivalent, or if
two infinite processes are testing equivalent up to a certain depth.
Moreover, the testing semantics provides a way to compute the
testing outcomes in practice, without requiring access to the
internal structure of the system other than the actions-candidates
for synchronization between the system and the test. To our
knowledge, this is the first equivalence that accomplishes all of
the above stated goals.

\paragraph{Acknowledgements} We are  grateful to Jos Baeten,  Erik de Vink, Manuel
N\'{u}\~{n}ez, and the anonymous reviewers for their valuable
comments on a draft version of this paper.